\documentclass[pra,aps,twocolumn,superscriptaddress,showpacs,eqsecnum,nofootinbib]{revtex4}

\usepackage{amsmath}
\usepackage{amssymb}
\usepackage{graphicx}
\usepackage{color}
\usepackage{hyperref}
\usepackage{ulem}
\usepackage[notrig]{physics}
\usepackage{mathtools}
\usepackage{MnSymbol}
\usepackage{relsize}

\begin{document}
\title{Spatial dependence of fidelity for a two-qubit Rydberg-blockade quantum gate}

\author{I. Vybornyi}
\affiliation{Institut f\"ur Theoretische Physik, Leibniz Universit\"at Hannover, Appelstrass e 2, 30167 Hannover, Germany}
\author{L.V. Gerasimov}
\affiliation{Quantum Technology Centre, Faculty of Physics, M.V. Lomonosov Moscow State University,\\ Leninskiye Gory 1-35, 119991, Moscow, Russia}
\affiliation{Center for Advanced Studies, Peter the Great St. Petersburg Polytechnic University, 195251, St. Petersburg, Russia}
\author{D.V. Kupriyanov}
\affiliation{Quantum Technology Centre, Faculty of Physics, M.V. Lomonosov Moscow State University,\\ Leninskiye Gory 1-35, 119991, Moscow, Russia}
\affiliation{Department of Physics, Old Dominion University 4600 Elkhorn Ave. Norfolk, VA 23529 USA}
\author{S.S. Straupe}
\affiliation{Quantum Technology Centre, Faculty of Physics, M.V. Lomonosov Moscow State University,\\ Leninskiye Gory 1-35, 119991, Moscow, Russia}
\affiliation{Russian Quantum Center, Skolkovo, Moscow 143025, Russia}
\author{K.S. Tikhonov}
\affiliation{St. Petersburg State University, 199034, St. Petersburg, Russia}
\affiliation{Russian Quantum Center, Skolkovo, Moscow 143025, Russia}
\begin{abstract}

We study the spatial performance of the entangling gate proposed by H. Levine et al. (Phys. Rev. Lett. 123, 170503 (2019)). This gate is based on a Rydberg blockade technique and consists of just two global
laser pulses which drive nearby atoms. We analyze the multilevel Zeeman structure of interacting $^{87}$Rb Rydberg atoms and model two experimentally available excitation schemes using specific driving beams geometry and polarization. In particular, we estimate the blockade shift dependence on inter-atomic distance and angle with respect to the quantization axis. In addition, we show that using Rydberg $d$-states, in contrast to $s$-states, leads to a pronounced angular dependence of the blockade shift and gate fidelity.
\end{abstract}
\pacs{pacs}
\maketitle

\section{Introduction}\label{Section_I}

One of the biggest challenges of modern atomic physics and quantum optics is an implementation of quantum computing at the physical level \cite{Henriet_2020}. An ensemble of trapped neutral atoms is an attractive platform for large quantum information systems, which demonstrates high potential for scalability and can easily support up to hundreds of qubits in a single array \cite{Ebadi2021QuantumPO,Wu_2019,PhysRevLett.123.230501}. The developed trapping techniques allow for precise individual control and manipulation at the single-qubit level as well as high isolation from the environment. 

For decades, the main bottleneck of this platform has been the lack of robust and efficient techniques for implementing multi-qubit entangling gates. The existing protocols \cite{PhysRevLett.82.1060,PhysRevLett.82.1975,PhysRevA.62.052302} with Rydberg-atom based techniques typically being the tool of choice \cite{Shi_2022} still suffer from low operational fidelity, which does not allow one to implement large-scale fault-tolerant quantum computing and which is mostly caused by imperfect coherent control of ground-Rydberg excitation \cite{Henriet_2020}. Furthermore, the overwhelming complexity of Rydberg atoms short-range interaction may manifest itself in undesired effects potentially negating the protocol efficiency \cite{Dereviyanko2015,Deutsch2013}. However, recent advances in Rydberg atom control \cite{PhysRevA.97.053803,PhysRevLett.121.123603} opened new opportunities for realization of entangling gates, in particular resulting in a new approach to the controlled-phase gate \cite{Lukin2019}. This gate is based on a novel protocol consisting of two global laser pulses which drive nearby atoms within the Rydberg blockade regime and experimentally reaches fidelity up to $97.4(3)\%$. This protocol is also experimentally convenient since only global pulses are required, thus eliminating the requirement for fast switching of the addressing beams between the atoms during the protocol. 

Since the gate operation was considered in \cite{Lukin2019} only for one-dimensional arrays of atoms, the next natural step towards parallel multiqubit quantum computing would be to generalize it to the case of two-dimensional (or even three-dimensional) optical lattices. For this purpose, in this article we analyze the spatial performance of the gate based on a particular choice of the Rydberg states of $^{87}$Rb atoms with different angular momentum quantum numbers and inter-atomic distances. In contrast to \cite{Browaeys2014,Browaeys2015}, where the general angular dependence of several interacting Rydberg atoms was studied, our goal will be to find a state that is optimal in terms of the two-qubit entangling CZ gate fidelity. We also analyze the multilevel Zeeman structure of interacting Rydberg atoms and model two experimentally available excitation schemes using specific driving beams geometry and polarizations to obtain strong van der Waals interactions allowing for single-atom excitation only. The obtained results  will be a helpful reference for an effective choice of proper geometry and will be of particular interest for quantum computing experiments with Rydberg atoms.

The remainder of the article is organized as follows. In Sec.~\ref{Section_IIA}  we provide a brief description of the controlled-phase gate of interest and consider basic requirements for its implementation. In Sec.~\ref{Section_IIB} we discuss how the van der Waals interaction between two Rydberg atoms depends on interatomic separations and angular momentum channels. Based on this analysis in Sec.~\ref{Section_IIС} we estimate the blockade shift dependence on the interatomic distance and angle with respect to the quantization axis. Finally, in Sec.  \ref{Section_III} we calculate fidelity of the controlled-phase gate for different spatial configurations.


\section{Two-qubit Rydberg CZ gate} \label{Section_II}
\subsection{Protocol description}\label{Section_IIA}

\begin{figure}[tp]
\includegraphics[width=8.6cm]{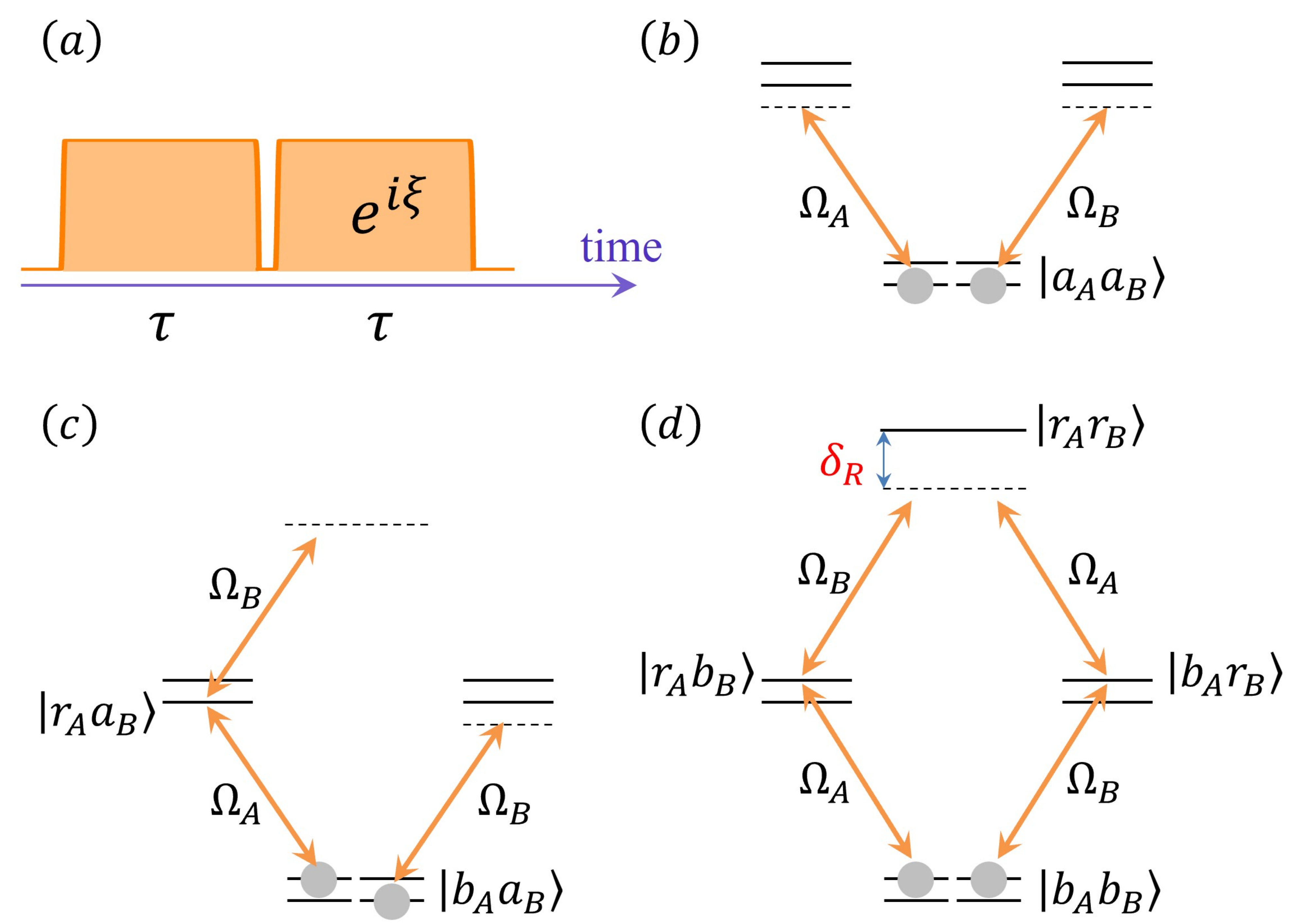}
\caption{(a)  Sequence of two global pulses providing coupling between qubit $|b\rangle$-states and Rydberg $|r\rangle$ states of atoms $A$ and $B$. The driving fields' Rabi frequencies are $\Omega_A = \Omega_B$ and $\xi$ is the respective phase shift between the pulses. 
(b) If both atoms are in the lower spin states $|a\rangle$ the pulse sequence does not change their collective state, and the phase shift $\phi_{aa}$ is accumulated only due to the Stark effect.
(c) If atom $A$ occupies a Zeeman state $|b\rangle$ belonging to the upper hyperfine sublevel and atom $B$ is in a state $|a\rangle$ belonging to the lower sublevel, the pulse sequence coupled with the Rydberg state $|r\rangle$ initiates a two-level transition between collective states $|b,a\rangle$ and $|r,a\rangle$ with a Rabi frequency $\Omega_A$. The evolution of the system from the initial state $|a,b\rangle$ is symmetric due to equivalence of the control fields acting on the atoms. The phase shift $\xi$ is chosen to guarantee that the system population returns to the initial state.
(d) If both atoms occupy the state $|b\rangle$, their excitation by the pulses to states $|r\rangle$ shifts the energy level by a value $\delta_R$ and eliminates the coupling to the state $|r,r\rangle$ in the Rydberg blockade regime. In this case, the effective Rabi frequency is enhanced by a factor $\sqrt{2}$ (see text for details). 
}
\label{Fig1}%
\end{figure}%

\noindent In this section we provide a brief description of the controlled-phase protocol for the two-qubit entangling CZ-gate originally proposed in \cite{Lukin2019}. One of the key advantages of this protocol is that it implies the control pulses driving both atoms simultaneously, which follows the idea of global operations in trapped ion systems \cite{Sorensen1999}. The qubits are encoded into the hyperfine clock transition of alkali-metal atoms $A$ and $B$, where the lower (logical ``0'') and the upper (logical ``1'') spin sublevels are denoted as $|a\rangle_{A,B}$ and $|b\rangle_{A,B}$ respectively. We assume that the two-photon excitation process provides coupling of the qubit states $|b\rangle_{A,B}$ to the Rydberg sublevels $|r\rangle_{A,B}$ with effective Rabi frequencies $\Omega_A = \Omega_B = \Omega$ and two-photon detunings $\Delta_A = \Delta_B = \Delta$. The sequence of two control pulses with a relative phase shift $\xi$ is shown in Fig.~\ref{Fig1}(a). Figs.~\ref{Fig1}(b-d) illustrate how the different basis states are transformed. A strategic goal of the interaction process is to ensure the necessary phases in the output state and the return of qubit's occupations to the initial states. A proper adjustment of the experimental parameters, such as the pulse duration $\tau$, phase $\xi$ and the $\Delta/\Omega$ ratio, allows one to calibrate the performed unitary transformation for an implementation of the desired CZ-gate. Let us assume that we originally prepare our system in the following state:
\begin{equation}
|\psi\rangle_A \otimes |\psi\rangle_B = \frac{1}{\sqrt{2}}[|a\rangle + |b\rangle]_A \otimes \frac{1}{\sqrt{2}}[|a\rangle + |b\rangle]_B, 
    \label{2.1}
\end{equation}
where we make use of the interaction representation with respect to the internal degrees of freedom of both atoms. Its transformation by the control pulse sequence results in the following output state
\begin{eqnarray}
&&|\psi\rangle_{AB} = \nonumber\\ && \frac{1}{2}[{\rm e}^{i\phi_{aa}}|a,a\rangle + {\rm e}^{i\phi_{ab}}|a,b\rangle + {\rm e}^{i\phi_{ba}}|b,a\rangle + {\rm e}^{i\phi_{bb}}|b,b\rangle]_{AB}\nonumber\\
&\propto&
\frac{1}{2}[|a,a\rangle - {\rm e}^{i\phi}|a,b\rangle - {\rm e}^{i\phi}|b,a\rangle - {\rm e}^{2i\phi}|b,b\rangle]_{AB}
    \label{2.2}
\end{eqnarray}
where each of the collective basis states accumulates a controllable phase shift, and both qubits accumulate a global phase $\phi$ which can be compensated by an additional spin rotation in an experiment. Note that despite being not involved in the two-photon excitation process, the qubit states $|a\rangle$ accumulate a global phase shift $\varphi_a$ associated with the Stark shifts induced to the hyperfine sublevels by the driving lasers. Hence, the first basis state $|a,a\rangle$ acquires a phase shift $\phi_{aa} = 2\varphi_a$, see Fig.~\ref{Fig1}(b).

The basis state $|b,b\rangle$ experiences an excitation by the two-photon pulses. In an ideal scenario of a perfect Rydberg blockade this process can be approximated by Rabi oscillations between $|b,b\rangle$ and  $[|r,b\rangle + |b,r\rangle]/\sqrt{2}$ states, parameterized by an enhanced Rabi frequency $\sqrt{2}\Omega$, see Fig.~\ref{Fig1}(d). The protocol implies that a complete Rabi oscillation cycle is performed during the first ($2\pi$)-pulse period $\tau$. This requires one to choose the pulse duration as $\tau = 2\pi/\sqrt{2|\Omega|^2 + \Delta^2}$ and makes the acquired phase $\phi_{bb}$ insensitive to the laser phase jump $\xi$, such that $\phi_{bb} = \Delta\tau$.

The two-photon process couples the basis states $|a,b\rangle$ and $|b,a\rangle$ to states $|a,r\rangle$ and $|r,a\rangle$ respectively and initiates the two-level dynamics with the corresponding Rabi frequency $\Omega$, see Fig.~\ref{Fig1}(c). The laser phase shift can be chosen to minimize leakage to the excited states, such that the control sequence returns the atoms to the qubit states $|b\rangle$, and the phase shifts $\phi_{ab} = \phi_{ba} = \Delta\tau - \xi + \varphi_a + \pi$ are acquired.

The correspondence between the last two lines in (\ref{2.2}) can be achieved by adjusting the ratio between the two-photon detuning and the effective Rabi frequency $\Delta/\Omega$, which also defines a global phase shift $\phi$. The latter can be compensated via a global spin rotation to convert the CZ transformation to its conventional form: ${\rm diag}(1,-1,-1,-1)$.

Note that the realization of the protocol implies that the qubit states $|b\rangle$ of both atoms are coupled to the corresponding Rydberg states $|r\rangle$ via an effective two-level transition scheme, that guarantees the required Rabi oscillations dynamics of the atomic population between the coupled states. Another necessary condition is a strong Rydberg blockade regime. Both of these requirements can be fulfilled by proper selection of an excitation scheme and a particular Rydberg state $|r\rangle$, which we discuss in the section below.


\subsection{Dipole interaction in the Rydberg states}
\label{Section_IIB}
\noindent 
To estimate the effective level shift $\delta_R$ preventing the system from entering the doubly-excited state (see Fig.~\ref{Fig1}(d)) and necessary for further fidelity estimation, one needs to consider in detail the Rydberg-Rydberg dipole interaction giving rise to the shift. In a situation where the two Rydberg atoms are separated well enough, the description of the arising dipole interaction is effectively split into different angular structure channels of the form
\begin{align}
nlj+nlj\rightarrow n_Al_Aj_A+n_Bl_Bj_B,
\label{channel}
\end{align}
where $n,l,j$ is the set of quantum numbers describing the initially populated Rydberg state and $n_{A,B}, l_{A,B}, j_{A,B}$
describe the intermediately populated two-atom states. For Rydberg interactions, the hyperfine splitting does not play a major role and thus this degree of freedom is omitted. Typically, several channels simultaneously contribute to the long-range interaction of the two atoms and the overall Hamiltonian is written as a sum of van der Waals type Hamiltonians over the relevant individual channels (\ref{channel}):
\begin{align}
    {\hat H}_{\rm vdW}=\sum_{k}\frac{C_6^{(k)}}{R^6} \mathcal{{\hat D}}_k
    \label{eq:ham}
\end{align}
with $\mathcal{{\hat D}}_k$ being dyad-type operators reproducing angular dependence of the interaction for a particular $k$-th channel, thoroughly described in \cite{Walker_2008}, $R$ being the interatomic distance\footnote{Operators $\mathcal{{\hat D}}_k$ can be naturally linked with a set of irreducible tensor operators, defined for each atom in the basis of the angular momentum states belonging to a particular atomic multiplet, see \cite{HapperReview,OmontReview,Varshalovich}}. For a particular index $k$ defining the angular structure of the interaction (i.e. the indices $l_A,j_A,l_B,j_B$), the $C_6^{(k)}$ coefficient incorporates the dependence on the atomic energy level structure in the vicinity of the considered Rydberg state and is given by the following sum over the principle numbers of the coupled channels:
\begin{align}
    C_6^{(k)} = \sum_{n_A,n_B}\frac{c^2\alpha^2}{-\delta_{AB}}\qty(R_{nl}^{n_Al_A}R_{nl}^{n_Bl_B})^2,
    \label{2.4}
\end{align}
where $\alpha$ is the fine-structure constant and we set $\hbar = 1$.
In practice, only a small number of states make a considerable contribution to the van der Waals interaction and for every particular channel $k$ the sum is reduced to a certain subset of nearby-lying Rydberg states with smallest energy defects
\begin{equation}
\delta_{AB} = E_{n_Al_Aj_A}+E_{n_Bl_Bj_B}-2E_{nlj},\label{energy_defects}
\end{equation}
where the energy of highly excited state of an alkali-metal atom with nearly hydrogenic potential is given by
\begin{equation}
E_{nlj} = - \frac{1}{2}\,m_{\rm e}c^2 \frac{\alpha^2}{(n + \Delta_l)^2}
 - \frac{1}{2}\,m_{\rm e}c^2 \frac{\alpha^4}{n^4}\left(\frac{n}{j+\frac{1}{2}} - \frac{3}{4}\right) + \ldots,
    \label{2.5}
\end{equation}
where $m_{\rm e}$ is the electron mass and the omitted higher relativistic corrections are denoted by ellipses \cite{messiah61,LaLfIII}. An angular momentum dependent quantum defect $\Delta_l$ accounts for the corrections to the Coulomb potential by the core electrons and vanishes with increasing $l$. The number of terms to include in the sum (\ref{2.4}) varies based on the chosen initial state. Generally, states with higher angular momentum produce more relevant terms in the sum as the energy defects of the nearby states are smaller. The radial matrix elements are given by
\begin{equation}
R_{nl}^{n'l'} = \langle n'l' | r | nl \rangle =\int r \chi^\ast_{n'l'}(r)\chi_{nl}(r) dr,
\end{equation}
where $\chi_{nl}$ are the corresponding radial atomic wavefunctions in the effective potential. 

The Hamiltonian (\ref{eq:ham}) acts within the Hilbert subspace spanned by the degenerate Zeeman sublevels $\ket{m_A}\otimes \ket{m_B}\equiv\ket{m_Am_B}$ of the initial two-atom Rydberg state, where $m_A$ and $m_B$ are the projections of the total angular momenta $\mathbf{j}_A$ and $\mathbf{j}_B$ on the quantization axis. It produces a set of eigenvectors $\ket{\varphi}$ and eigenenergies $\Delta E_{\varphi}$, which are necessary for blockade shift calculations: 
\begin{align}
    {\hat H}_{\rm vdW}\ket{\varphi}=\Delta E_{\varphi}\ket{\varphi}.
\end{align}
The choice of the van der Waals form for the Hamiltonian \eqref{eq:ham} will be justified further by the calculation of the resulting Rydberg defects given by (\ref{energy_defects}) which are an order of magnitude larger then the corresponding blockade shifts (see Fig. \ref{Fig5}).

The Rydberg states we consider are $|70s(^2S_{1/2})\rangle$ and $|70d(^2D_{3/2})\rangle$ for $^{87}$Rb. The choice corresponds to typical values of $n$ used in experiments and is a reasonable balance between larger values of the blockade shift and higher polarizability giving rise to increased sensitivity to external electric fields.

For $|70s(^2S_{1/2})\rangle$, the contributing angular structure channels (\ref{channel}) are:
\begin{align}
    1. s_{1/2}+s_{1/2} &\rightarrow p_{1/2}+p_{1/2},\nonumber\\
    2. s_{1/2}+s_{1/2} &\rightarrow p_{1/2}+p_{3/2},\nonumber\\
    3. s_{1/2}+s_{1/2} &\rightarrow p_{3/2}+p_{3/2}.\nonumber\\
    \label{channels70s}
\end{align}
The sum in $C^{(k)}_6$ coefficients to a good approximation could be reduced to one or two terms only and for each channel evaluates to $C_6^{(1)}=437$ GHz$\cdot\mu$m$^6$, $C_6^{(2)}=1132$ GHz$\cdot\mu$m$^6$, $C_6^{(3)}=799$ GHz$\cdot\mu$m$^6$. This allows us to compute the sought-for spectrum of the interaction Hamiltonian (\ref{eq:ham}).

For $|70d(^2D_{3/2})\rangle$, more angular structure channels (\ref{channel}) contribute to the interaction:
\begin{align}
    1. d_{3/2}+d_{3/2} &\rightarrow p_{3/2}+p_{3/2},\nonumber\\
    2. d_{3/2}+d_{3/2} &\rightarrow p_{3/2}+p_{1/2},\nonumber\\
    3. d_{3/2}+d_{3/2} &\rightarrow p_{1/2}+p_{1/2},\nonumber\\
    4. d_{3/2}+d_{3/2} &\rightarrow p_{3/2}+f_{5/2},\nonumber\\
    5. d_{3/2}+d_{3/2} &\rightarrow p_{1/2}+f_{5/2},\nonumber\\
    6. d_{3/2}+d_{3/2} &\rightarrow f_{5/2}+f_{5/2}.\nonumber\\
    \label{channels70d}
\end{align}
The sum in $C_6$ coefficients is reduced to two or three closest states and evaluates to $C_6^{(1)}=68$ GHz$\cdot\mu$m$^6$, $C_6^{(2)}=57$ GHz$\cdot\mu$m$^6$, $C_6^{(3)}=57$ GHz$\cdot\mu$m$^6$, $C_6^{(4)}=-2827$ GHz$\cdot\mu$m$^6$, $C_6^{(5)}=-6070$ GHz$\cdot\mu$m$^6$, $C_6^{(6)}=-32$ GHz$\cdot\mu$m$^6$. 

\subsection{Blockade shift}
\label{Section_IIС}
\begin{figure}[tp]
\includegraphics[width=8.6cm]{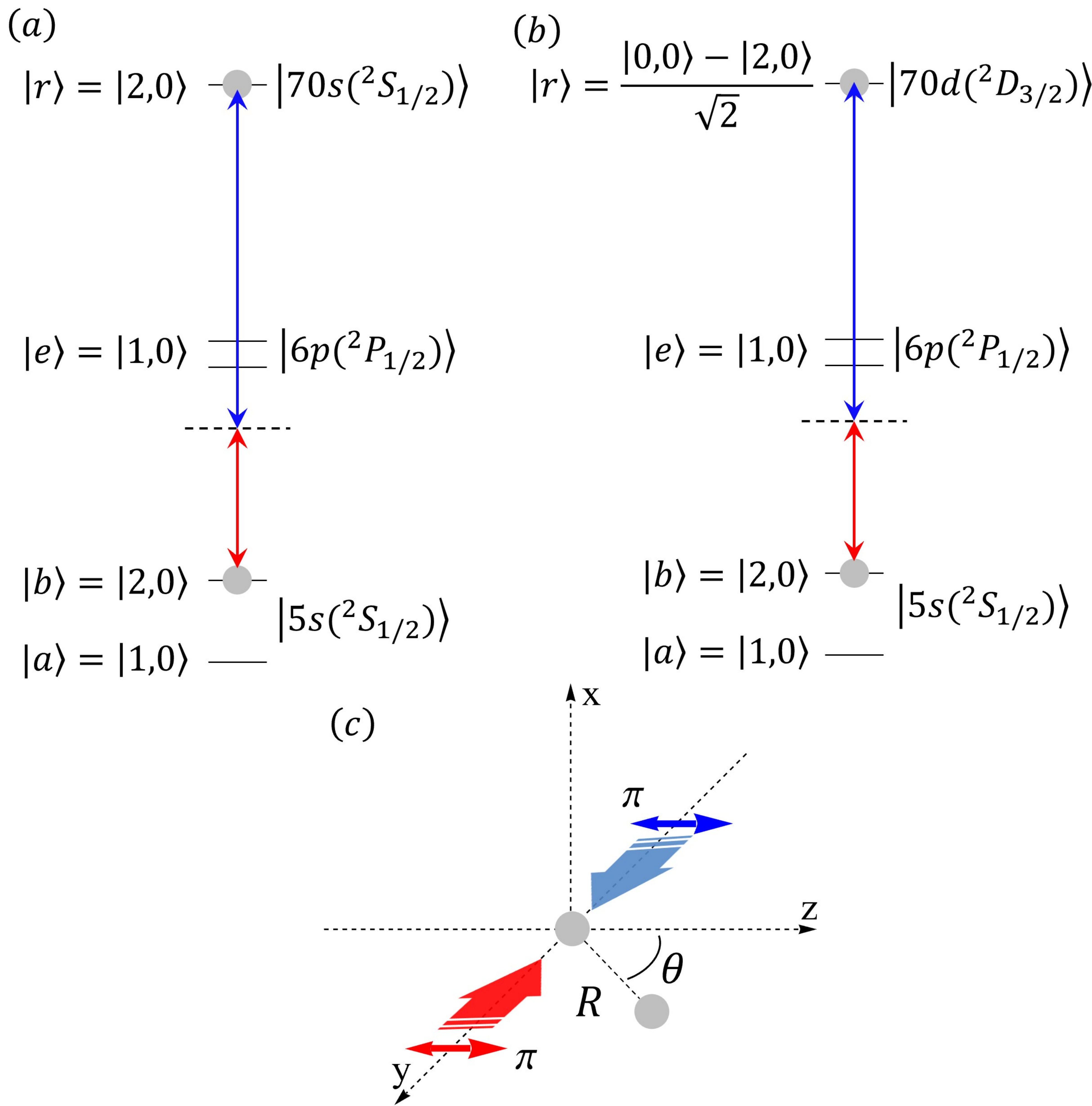}
\caption{The transition diagrams for $^{87}$Rb driven by two counter-propagating and linearly polarized light beams. The participating states are specified by the concrete numbers of the total electronic and nuclear spin angular momenta and their projections. The used energy configuration of $^{87}$Rb is effectively two-level and provides coupling of the qubit state $|b\rangle$ only with a single Rydberg state $|r\rangle$ having the principal quantum number $n = 70$. We consider excitation to the Rydberg states with two different combinations of orbital and total angular momenta: (a) $l_r = 0$, $j_r = 1/2$, when only one excited sublevel $|r\rangle = |F_r=2,M_r=0\rangle$ is involved in the excitation process; (b) $l_r = 2$, $j_r = 3/2$, when two Rydberg sublevels $|F_r=0,M_r=0\rangle$ and $|F_r=2,M_r=0\rangle$ are coupled to the qubit $|b\rangle$ state, and form the coherent superpositions of mutually-orthogonal ``bright'' (involved into the two-photon interaction process) and ``dark'' (isolated from the coherent coupling) states. (c) Excitation geometry: driving fields are linearly polarized along the $z$-axis. We consider the blockade shift as a function of the interatomic distance $R$ and the angle between the molecular axis and the $z$-axis  $\theta$.}
\label{Fig2}%
\end{figure}

\noindent
The coupling to the Rydberg state of choice is supposed to proceed via a two-photon process allowed by the selection rules. The gate operating scenario shown in Fig.~\ref{Fig1}(d) involves a pair of atoms being pumped simultaneously into the Rydberg state. In this situation, the doubly-excited state of the pair appears to be shifted from the resonance by a value of the \textit{mean blockade shift} $\delta_R$, given by \cite{Walker_2008}:
\begin{align}
    \frac{1}{\delta_R^2}=\sum_{\varphi}\frac{\kappa_{\varphi}^2}{\Delta E_{\varphi}^2},
\end{align}
where the sum is taken over the spectrum of Rydberg interaction Hamiltonian (\ref{eq:ham}) with eigenenergies $\Delta E_{\varphi}$ and overlap factors $\kappa_{\varphi}=\braket{\varphi}{\Psi}$. Here $\ket{\varphi}$ is a particular interaction eigenstate and $\ket{\Psi}$ is the two-atom Zeeman state which we attempt to populate with the laser field. In order to determine the form of $\ket{\Psi}$ we need to specify the particular schemes of two-photon ground-Rydberg excitation. Considered geometries and transition diagrams are shown in Figs.~\ref{Fig2}(a,b).

In our calculations we consider two $^{87}$Rb atoms and specify the hyperfine qubit states as $|a\rangle = |5s(^2S_{1/2});F_0 = 1, M_0 = 0\rangle$ and $|b\rangle = |5s(^2S_{1/2});F_0 = 2, M_0 = 0\rangle$. The qubit state $|b\rangle$ is coupled to the considered Rydberg states $|r\rangle$ via the two-photon excitation process initiated by two counter-propagating beams linearly-polarized along $z$-axis as shown in Fig.~\ref{Fig2}(c). Note that only a single intermediate state $|e\rangle = |6p(^2P_{1/2});F = 1, M = 0\rangle$ contributes to the ladder-type two-photon excitation process in this excitation geometry due to convenient selection rules which allow us to minimize the losses due to incoherent scattering and ignore the negative effect of photon recoil.\footnote{We assume that the atoms are trapped in optical tweezers with the beam axis along the $z$ direction at all times excluding the course of the protocol. The counter-propagating driving beams push atoms in the transverse plane, where they have tighter confinement than in the axial direction and can be effectively frozen via a Raman sideband cooling protocol.}

Excitation to $|70s(^2S_{1/2})\rangle$ selects a single Zeeman state $|r\rangle = |F_r = 2, M_r = 0\rangle$, see Fig.~\ref{Fig2}(a). Thus $\ket{\Psi}$ has the following form in the basis of two-atom projections of angular momentum $\ket{m_A}\otimes \ket{m_B}$, where the projection of the nuclear spin is ignored:
\begin{eqnarray}
    \ket{\Psi_s}&=&\mbox{$\frac{1}{\sqrt{2}}$}\qty[\ket{\mbox{$+\frac{1}{2}$}}+\ket{\mbox{$-\frac{1}{2}$}}]_A\otimes\mbox{$\frac{1}{\sqrt{2}}$}\qty[\ket{\mbox{$+\frac{1}{2}$}}+\ket{\mbox{$-\frac{1}{2}$}}]_B\nonumber\\
    &=&\mbox{$\frac{1}{2}$}\qty(\ket{\mbox{$+\frac{1}{2},+\frac{1}{2}$}}+\ket{\mbox{$-\frac{1}{2},-\frac{1}{2}$}}+\ket{\mbox{$+\frac{1}{2},-\frac{1}{2}$}}+\ket{\mbox{$-\frac{1}{2},+\frac{1}{2}$}})\nonumber,\\
    \label{psi_s}
\end{eqnarray}
For the $|70d(^2D_{3/2})\rangle$ state two Rydberg sublevels are accessed via two-photon transitions: $|F_r=0,M_r=0\rangle$ and $|F_r=2,M_r=0\rangle$, see Fig.~\ref{Fig2}(b). Thus, a slightly more complicated `bright state' is selected in the excitation process $|r\rangle = [|0,0\rangle - |2,0\rangle]/\sqrt{2}$, while the orthogonal `dark state' $[|0,0\rangle + |2,0\rangle]/\sqrt{2}$ is insensitive to the coherent coupling. This leads to the following form of the double Rydberg state written in the uncoupled basis
\begin{eqnarray}
    \ket{\Psi_d}&=&\mbox{$\frac{1}{\sqrt{2}}$}\qty[\ket{\mbox{$+\frac{1}{2}$}}-\ket{\mbox{$-\frac{1}{2}$}}]_A\otimes\mbox{$\frac{1}{\sqrt{2}}$}\qty[\ket{\mbox{$+\frac{1}{2}$}}-\ket{\mbox{$-\frac{1}{2}$}}]_B\nonumber\\
    &=&\mbox{$\frac{1}{2}$}\qty(\ket{\mbox{$+\frac{1}{2},+\frac{1}{2}$}}+\ket{\mbox{$-\frac{1}{2},-\frac{1}{2}$}}-\ket{\mbox{$+\frac{1}{2},-\frac{1}{2}$}}-\ket{\mbox{$-\frac{1}{2},+\frac{1}{2}$}})\nonumber.\\
    \label{psi_d}
\end{eqnarray}

Note again, that the dependence on the nuclear spin degree of freedom is omitted in the above expressions, since the effects connected with hyperfine interactions do not play any significant role in estimating the blockade shift. However, they are of great importance when considering the Rydberg excitation geometry and are thus included in the further fidelity analysis as shown in the transition diagrams of Fig.~\ref{Fig2}(a,b).

The overlap factors $\kappa_{\varphi}$ are evaluated by switching to a common reference frame, which makes the factors dependent on the angle $\theta$ between the interatomic axis and the polarization of the driving field. A further comment on the observed angular dependency from a semiclassical prospective is given in the Appendix \ref{appx}.

The blockade shift $\delta_R$ is thus dependent on two spatial parameters: the interatomic distance $R$ and the angle $\theta$, see Fig.~\ref{Fig2}(c). For the two considered Rydberg states, the blockade shift spatial dependence is plotted in the upper panels of Fig.~\ref{Fig5},~\ref{Fig6}.

\begin{figure}[tp]
\includegraphics[width=8.6cm]{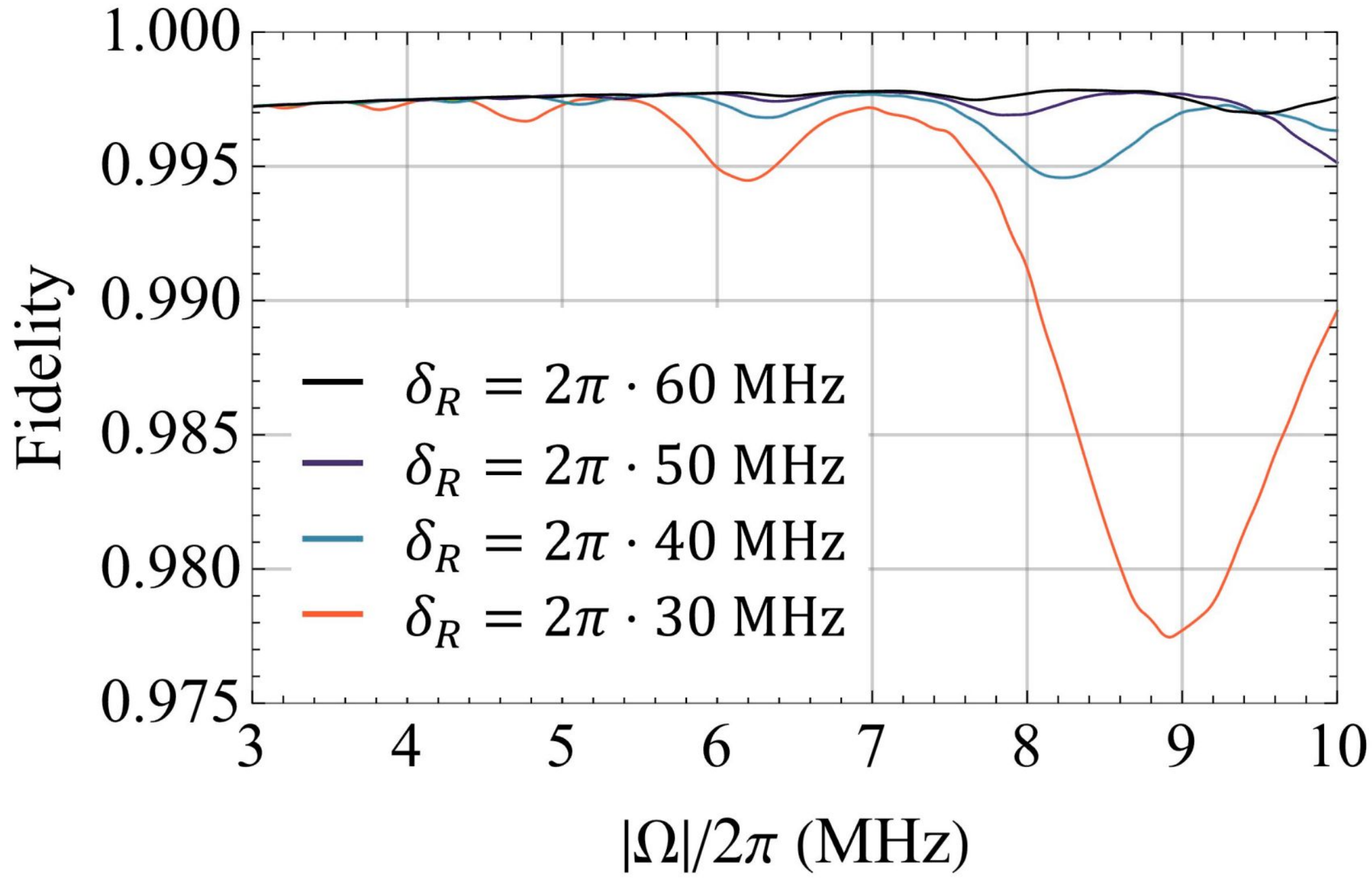}
\caption{Fidelity ${\cal F}$ of the entangled state prepared by a CZ transform via a controlled phase protocol as a function of the effective Rabi frequency $|\Omega|$ at varied values of the Rydberg blockade shift $\delta_R/2\pi = 30,\,40,\,50,\,60$ MHz. The oscillations indicate the probability of simultaneous occupation of $|r\rangle$ by atoms $A$ and $B$.}
\label{Fig3}%
\end{figure}%

\section{Results and discussion}\label{Section_III}
\noindent In this section we present the results of our numerical simulations for fidelity of the entanglement protocol. We analyze the results and compare the cases of different experimentally accessible excitation geometries to optimize the entanglement preparation. 

\begin{figure}[tp]
\includegraphics[width=8.6cm]{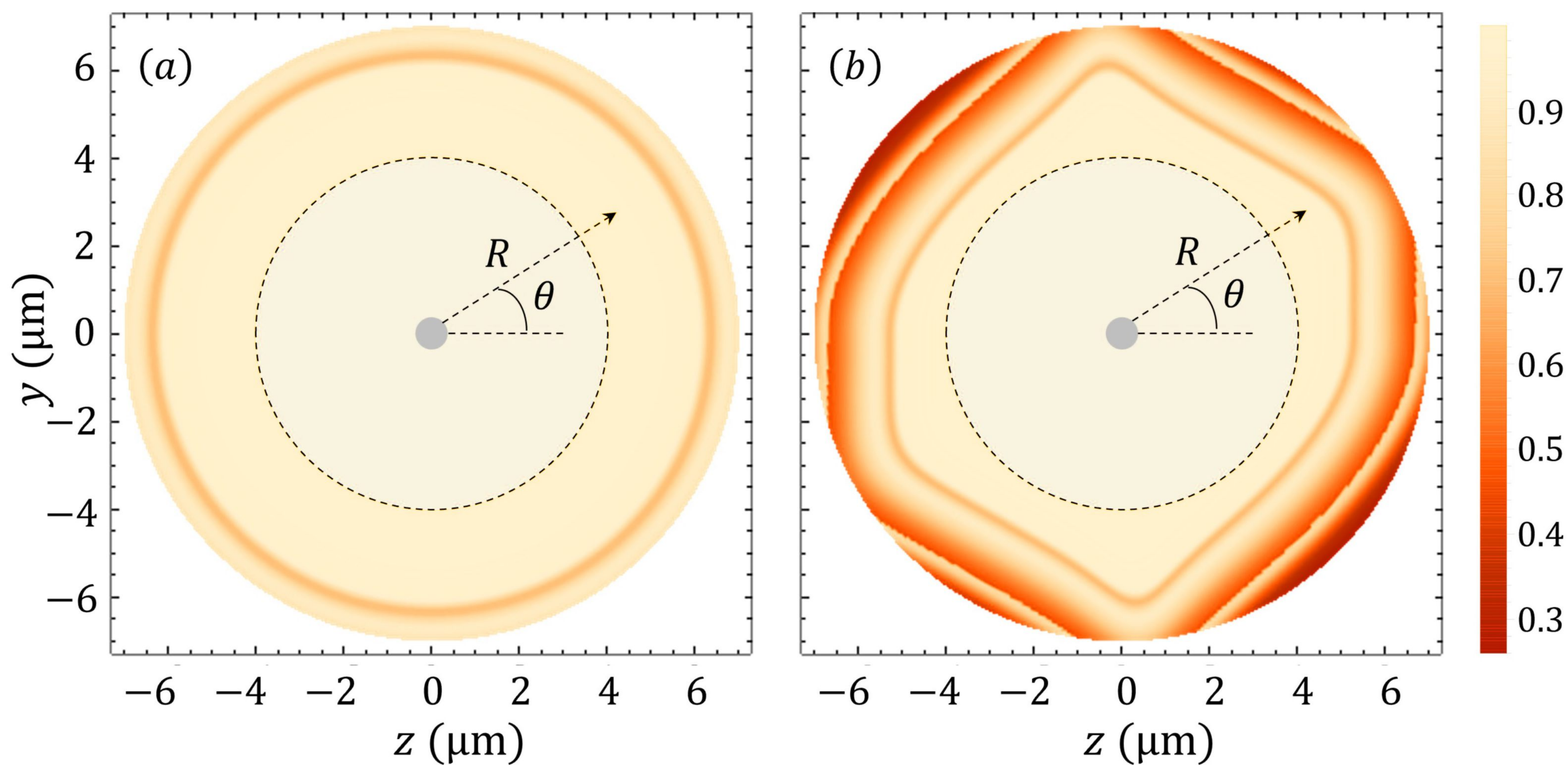}
\caption{CZ-gate fidelity ${\cal F} = {\cal F}(R,\theta)$ as a function of spatial and angular parameters $z = R\cos\theta$, $y = R\sin\theta$ for the excitation schemes shown in Fig.~\ref{Fig2}(a) (left panel) and  Fig.~\ref{Fig2}(b) (right panel). The shaded area corresponds to relatively short distances of several microns, where the interatomic potential deviates from $\sim R^{-6}$ scaling (see text for more details).}
\label{Fig4}%
\end{figure}%

\subsection{Fidelity analysis}


\noindent Fidelity of reproduction of an ideal output state $|\psi\rangle_{AB}$ defined by (\ref{2.2}) can be estimated as follows
\begin{equation}
{\cal F} = \langle\psi|\hat{\rho}|\psi\rangle_{AB}\, ,%
\label{3.1}%
\end{equation}
where $\hat{\rho}$ is the density operator of the mixed spin state incorporating incoherent processes associated with finite radiative lifetime of the Rydberg state $|r\rangle$ and the channels of incoherent Rayleigh and Raman scattering of the driving fields via intermediate states. It also accounts for the protocol imperfection due to the simultaneous occupation of the Rydberg state by both atoms $A$ and $B$. The problem of blockade leakage may become critical if the condition $|\Omega| \ll \delta_R$ is not fulfilled and here we mainly discuss the spatial variations of $\delta_R$ which subsequently lead to spatial dependence of the gate fidelity. Hence, we omit the description of the calculation scheme we use to derive the density operator of the output state, which will be published elsewhere \cite{TwoQubitsPaper2022}.

In Fig.~\ref{Fig3} we show the CZ-gate fidelity at varied and given values of blockade shifts $\delta_R/2\pi = 30,\,40,\,50,\,60$ MHz as a function of the effective Rabi frequency $\Omega$. If the effective Rabi frequency increases the probability amplitude of double excitation of the Rydberg state becomes non-negligible, which results in extra phase shifts in the prepared entangled state and its deviations from the ideal state $|\psi\rangle_{AB}$. The probability of double occupation oscillates with frequency and amplitude depending on the ratio $\delta_R/|\Omega|$ which is shown in Fig.~\ref{Fig3}. In the opposite limit of small $\Omega$ and long excitation pulses $\tau$ the entanglement fidelity is reduced by the irreversible incoherent processes. In our further calculations we set $|\Omega|/2\pi \simeq 7$ MHz, which corresponds to the CZ-gate time of $2\tau \simeq 180$ ns and provides the entanglement fidelity of ${\cal F} \simeq 0.997$ in the perfect blockade limit. The dependence of fidelity on the lifetime of the considered Rydberg state $|70s(^2S_{1/2})\rangle$ or $|70d(^2D_{3/2})\rangle$ at selected blockade shifts $\delta_R$ is weak and unresolved in Fig.~\ref{Fig3} within the plot scale for the tested calculation domain.

Density plots in Figs.~\ref{Fig4}(a,b) visualize the calculated fidelity as a function of two spatial arguments ${\cal F} = {\cal F}(R,\theta)$ for the two considered excitation schemes. We note that using the $|70s(^2S_{1/2})\rangle$ state one obtains an almost isotropic interaction between the two atoms, which results in an isotropic behaviour of the blockade shift and CZ-gate fidelity (left panel). The strong angular dependence arises for the case of excitation via $|70d(^2D_{3/2})\rangle$, that may provide a useful tool for selective addressing of atoms in the qubit array (right panel). Below we discuss the dependence of fidelity on the interatomic separation and its variation with angle in more detail.

\begin{figure}[tp]
\includegraphics[width=8.6cm]{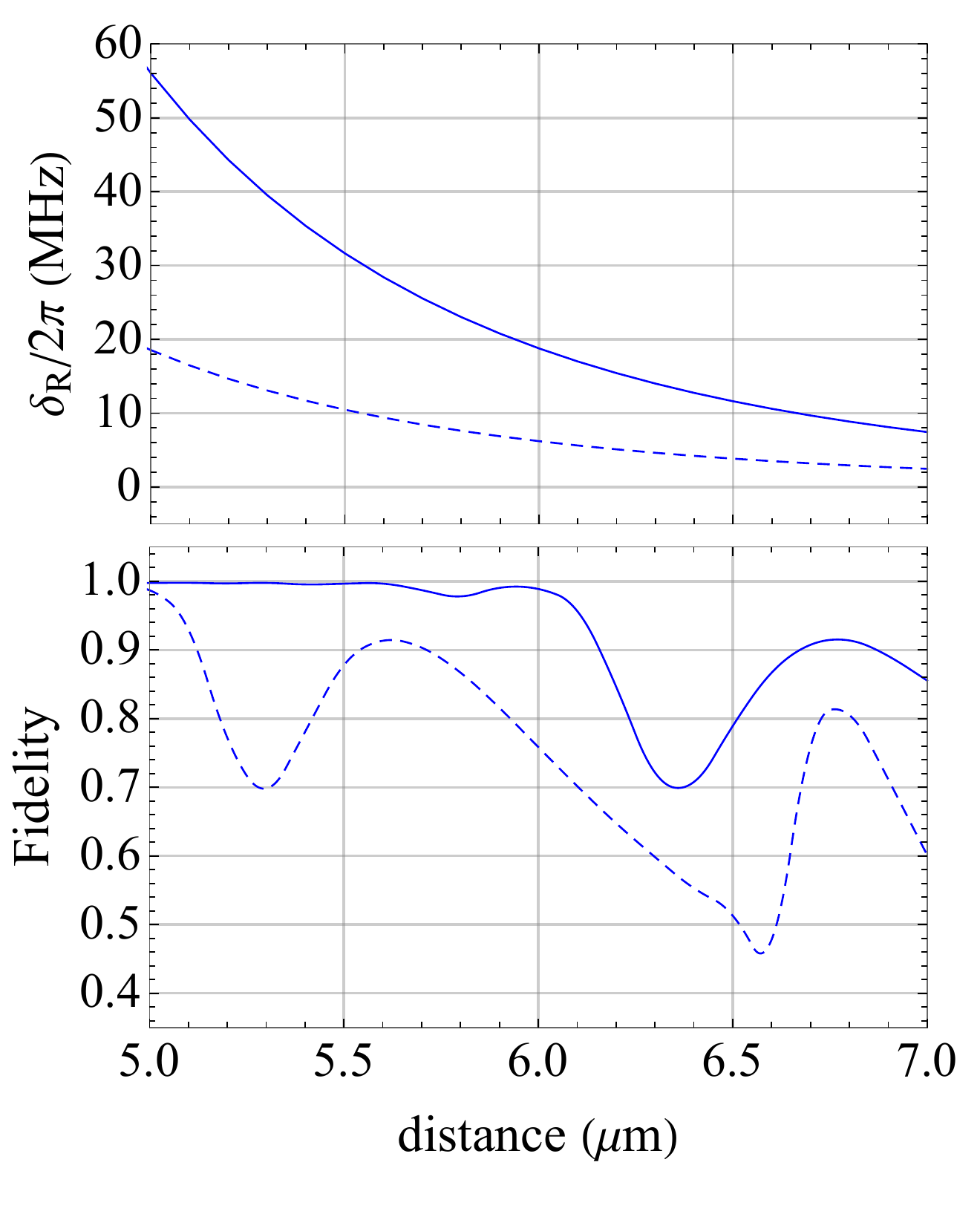}
\caption{Blockade shift (upper panel) and CZ-gate fidelity (lower panel) for excitation schemes via Rydberg states $|70s(^2S_{1/2})\rangle$ (solid curves) and $|70d(^2D_{3/2})\rangle$ (dashed curves) in $^{87}$Rb as a function of interatomic separation at angle $\theta=0$, see Figs.~\ref{Fig3}(c),\ref{Fig4}. The energy defects $\delta _{AB}$  for channels \eqref{channels70s} of $|70s(^2S_{1/2})\rangle$ and channels \eqref{channels70d} of $|70d(^2D_{3/2})\rangle$ of $10^2-10^3$ MHz are at least an order of magnitude larger then the blockade shift. }
\label{Fig5}%
\end{figure}%

The parameters $\delta_R$ and ${\cal F}$, calculated as a function of interatomic separation $R$, are shown in Fig.\ref{Fig5}. Here we specify the angle between the light polarization and the molecular axis $\theta = 0$ and observe the dependence of $\delta_R \propto R^{-6}$ (upper panel) which is typical for the Rydberg blockade regime. In contrast, fidelity ${\cal F} = {\cal F}(R)$ (lower panel) shows a non-monotonic behaviour for higher separation distances caused by its non-trivial oscillatory behaviour if the condition $|\Omega| \ll \delta_R$ is violated. The results are presented for the two considered Rydberg excitation schemes: $|70s(^2S_{1/2})\rangle$ (solid curve) and $|70d(^2D_{3/2})\rangle$ (dashed curve).

\begin{figure}[tp]
\includegraphics[width=8.6cm]{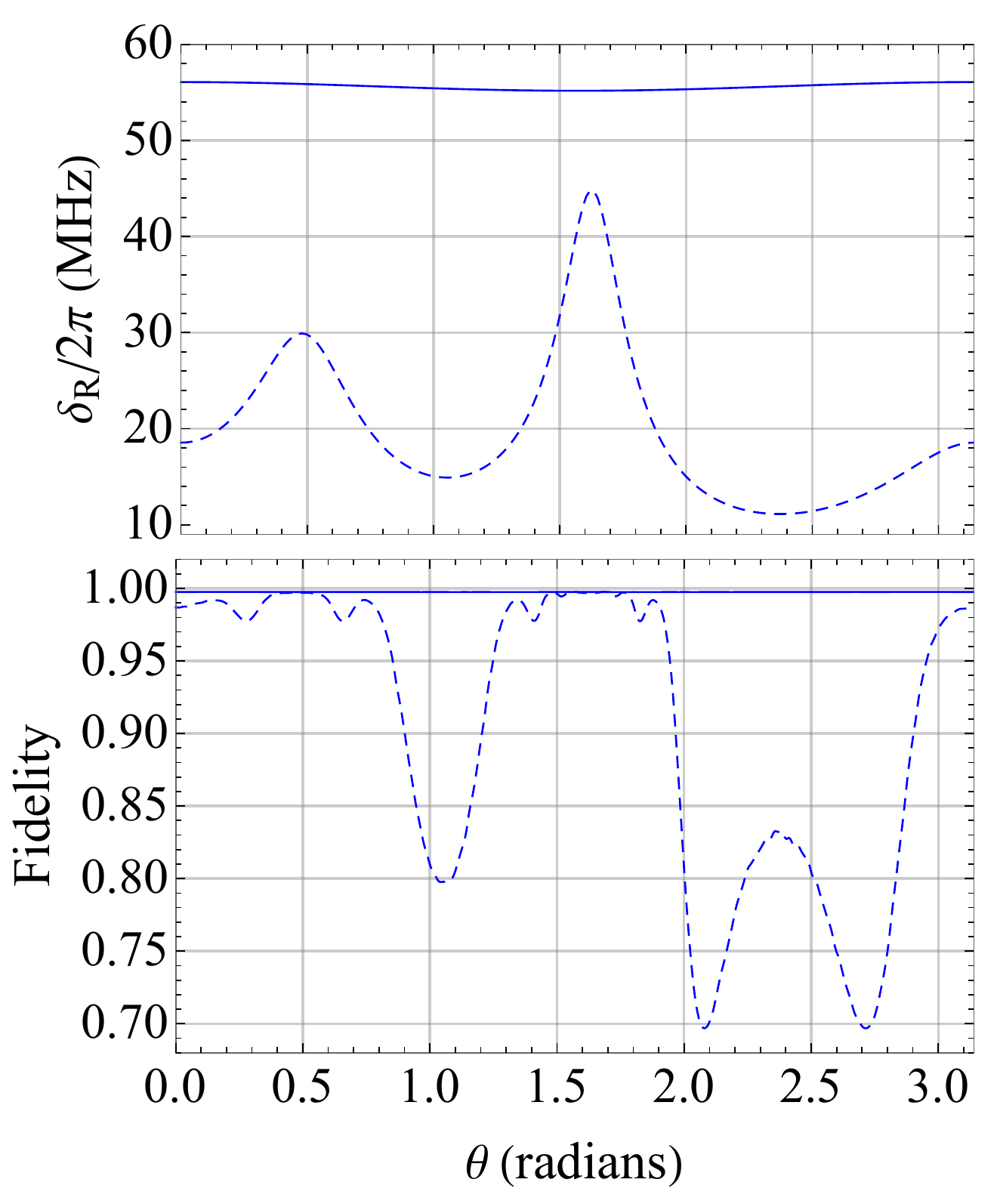}
\caption{Angular dependence of the blockade shift (upper panel) and CZ-gate fidelity (lower panel) for the excitation schemes via Rydberg states $|70s(^2S_{1/2})\rangle$ (solid curves) and $|70d(^2D_{3/2})\rangle$ (dashed curves) in $^{87}$Rb at the interatomic separation $R = 5\,\mu$m.}
\label{Fig6}%
\end{figure}%

The angular dependence of the CZ-gate fidelity ${\cal F} = {\cal F}(\theta)$ is presented in Fig.~\ref{Fig6}. The calculations performed at the interatomic separation $R = 5\, \mu$m and for the two considered schemes of atomic excitation via coupling of the qubit state $|b\rangle$ to the Rydberg states belonging to state $|70s(^2S_{1/2})\rangle$ (solid curves) and $|70d(^2D_{3/2})\rangle$ (dashed curves). The first choice of Rydberg $s$-states provides a nearly isotropic interaction, hence the gate fidelity does not depend on the orientation of the atoms with respect to the driving beams polarization. The situation is different if Rydberg states belong to the $d$-orbital, which results in a strong angular dependence of the blockade shift with a global maximum at $\theta\simeq \pi/2$. Note that typically, a maximum of $\delta_R$ corresponds to a higher ${\cal F}$. However the lowest values of fidelity are not always reproduced by the minima of the blockade shift, which is a further consequence of the non-monotonic behaviour of fidelity in a weak blockade regime $|\Omega| \sim \delta_R$.

\section{Conclusions}
\noindent In this paper we have considered the spatial dependence of fidelity of the atomic spin entanglement protocol based on the Rydberg blockade technique. We have analyzed the multilevel Zeeman structure of interacting $^{87}$Rb Rydberg atoms and modelled two experimentally available two-photon excitation schemes using specific driving beams geometry and polarizations to obtain strong van der Waals interactions allowing only single-atom excitation.

The two considered excitation schemes implied addressing to Rydberg states belonging to $|70s(^2S_{1/2})\rangle$ and $|70d(^2D_{3/2})\rangle$ manifolds having significantly different angular structure. The nearly isotropic interaction in Rydberg $s$-states results in a lack of angular dependence observed for the phase-controlled CZ-gate fidelity. The use of the lower angular momenta also seems reasonable to attain the strong Rydberg blockade regime since it corresponds to higher values of energy defects and less fine structure channels contributing to the interaction. 

We have shown, that using Rydberg $d$-states, in contrast to $s$-states, leads to a pronounced angular dependence of the blockade shift and gate fidelity. That, in turn, may be useful for engineering the strongly anisotropic interaction of atoms in a two-dimensional lattice.


\acknowledgements

\noindent

S.S.S., D.V.K. and L.V.G. acknowledge support from the Russian Science Foundation under Grant No. 18-72-10039. This work (excluding the contribution of I.V.) was supported by Rosatom in the framework of the Roadmap for Quantum computing (Contract No. 868-1.3-15/15-2021 dated October 5, 2021 and Contract No. P2154 dated November 24, 2021). S.S.S. acknowledges support by the Interdisciplinary Scientific and Educational School of Moscow University Photonic and Quantum Technologies, Digital Medicine.

\appendix
\section{Electronic density distribution of Rydberg $s$ and $d$-orbitals}\label{appx}
\noindent The qualitative behavior of highly excited Rydberg states in the Coulomb potential can be naturally described using a semiclassical visualization of the excitation process. Generally, the excited electron moves along an elliptical trajectory with its action variable quantized according to the Bohr-Sommerfeld principle. The anisotropy of the trajectory and its orientation can be incorporated into a single vector parameter, namely, the Runge–Lenz vector
\begin{equation}
\mathbf{A}=\mathbf{p}\times\mathbf{l}-m_{\rm e}e^2\mathbf{r}/{r}=\mathbf{n}\varepsilon,
\end{equation}
where $\mathbf{n}=\mathbf{r}/r$ and $\varepsilon$ are respectively the orbit director and its eccentricity. 

If the reference atom is excited to an $s$-orbital with zero orbital momentum $l=0$ then intuitively and naively we could expect that its classical trajectory and the Runge–Lenz vector would reduce to a line segment and to its unit director, respectively. However that is not so and such a comet-type trajectory cannot be simply reproduced by an $s$-orbital, since the former has an asymmetric charge distribution while the orbital wave-function does not. Specifics of the electron motion in a hydrogen-like Coulomb potential is that it admits arbitrary superpositions of the states of same energy but different $l$’s (different parities). In parabolic coordinates such a superposition can reveal the new set of states with $A_z\neq 0$, see \cite{LaLfIII}. Then the principal quantum number is given by
\begin{equation}
n=n_1+n_2+|m|+1,
\end{equation}
and the diagonal matrix elements of $A_z$ are
\begin{equation}
A_z=\frac{n_1-n_2}{n},
\end{equation}
where $n_1$ and $n_2$ are the parabolic quantum numbers and $|m|$ is the modulus of the azimuthal quantum number. Once we assume $m=0$ and either $n_1=n-1,n_2=0$ or $n_1=0,n_2=n-1$ we arrive at the quantum state optimally approaching the classical orbit with a zero angular momentum. Surprisingly, many orbital quantum numbers $l$’s are needed to construct such a state. 

In a realistic scenario, the states with different $l$ have different energies and the eigenstate problem can be fairly resolved only in spherical coordinates. Then the complete set of the commuting operators specifies the targeted $s$-state by its principle quantum number $n\sim 100$, and by a zero projection $m=0$ onto an arbitrary quantization direction. Any deviations for $A_z\neq 0$ would conflict with the isotropic charge distribution. Thus, for a highly excited $s$-orbital one should imagine its electronic density as a random isotropic probability distribution of its radial coordinate, with most typical values lying on a sphere formed by the apogee points of the classical orbits. In a semiclassical vision the electron spends most of its time just near the sphere surface, and we have fulfilled both the constraints $\langle\mathbf{A}\rangle =0$ (dictated by quantum mechanics) and $\langle\mathbf{A}^2\rangle\to 1$ (suggested by classical description).

If the reference atom is excited to a $d$-orbital with $l=2$ the situation is somewhat different. From the classical point of view the optical excitation affects both the linear and the orbital angular momentum of the electron motion. The classical trajectory becomes elliptical with $\langle\mathbf{A}^2\rangle\lesssim 1$ and the excitation by linearly polarized light provides $m=0$ only for the quantization direction along the light polarization. In a classical visualization of motion the Runge-Lenz vector is expected to be directed along the light polarization as well. But the occupied orbital has an even probability density function giving rise to a constraint $\langle\mathbf{A}\rangle =0$, i.e. the orbit director should have a zero projection on any axis. So the electronic density has a radial distribution, scaled by the separation of the orbit focuses, and, similarly to the case of the $s$-orbital, the most typical values of the radial coordinate belong to a surface formed by the orbit apogee points.  Nevertheless, its angular distribution becomes anisotropic and depending on the polar angle, manifesting in an anisotropic Rydberg blockade effect studied here.

\bibliographystyle{apsrev4-1}
\bibliography{references}

\begin{thebibliography}{23}%
\makeatletter
\providecommand \@ifxundefined [1]{%
 \@ifx{#1\undefined}
}%
\providecommand \@ifnum [1]{%
 \ifnum #1\expandafter \@firstoftwo
 \else \expandafter \@secondoftwo
 \fi
}%
\providecommand \@ifx [1]{%
 \ifx #1\expandafter \@firstoftwo
 \else \expandafter \@secondoftwo
 \fi
}%
\providecommand \natexlab [1]{#1}%
\providecommand \enquote  [1]{``#1''}%
\providecommand \bibnamefont  [1]{#1}%
\providecommand \bibfnamefont [1]{#1}%
\providecommand \citenamefont [1]{#1}%
\providecommand \href@noop [0]{\@secondoftwo}%
\providecommand \href [0]{\begingroup \@sanitize@url \@href}%
\providecommand \@href[1]{\@@startlink{#1}\@@href}%
\providecommand \@@href[1]{\endgroup#1\@@endlink}%
\providecommand \@sanitize@url [0]{\catcode `\\12\catcode `\$12\catcode
  `\&12\catcode `\#12\catcode `\^12\catcode `\_12\catcode `\%12\relax}%
\providecommand \@@startlink[1]{}%
\providecommand \@@endlink[0]{}%
\providecommand \url  [0]{\begingroup\@sanitize@url \@url }%
\providecommand \@url [1]{\endgroup\@href {#1}{\urlprefix }}%
\providecommand \urlprefix  [0]{URL }%
\providecommand \Eprint [0]{\href }%
\providecommand \doibase [0]{http://dx.doi.org/}%
\providecommand \selectlanguage [0]{\@gobble}%
\providecommand \bibinfo  [0]{\@secondoftwo}%
\providecommand \bibfield  [0]{\@secondoftwo}%
\providecommand \translation [1]{[#1]}%
\providecommand \BibitemOpen [0]{}%
\providecommand \bibitemStop [0]{}%
\providecommand \bibitemNoStop [0]{.\EOS\space}%
\providecommand \EOS [0]{\spacefactor3000\relax}%
\providecommand \BibitemShut  [1]{\csname bibitem#1\endcsname}%
\let\auto@bib@innerbib\@empty
\bibitem [{\citenamefont {Henriet}\ \emph {et~al.}(2020)\citenamefont
  {Henriet}, \citenamefont {Beguin}, \citenamefont {Signoles}, \citenamefont
  {Lahaye}, \citenamefont {Browaeys}, \citenamefont {Reymond},\ and\
  \citenamefont {Jurczak}}]{Henriet_2020}%
  \BibitemOpen
  \bibfield  {author} {\bibinfo {author} {\bibfnamefont {L.}~\bibnamefont
  {Henriet}}, \bibinfo {author} {\bibfnamefont {L.}~\bibnamefont {Beguin}},
  \bibinfo {author} {\bibfnamefont {A.}~\bibnamefont {Signoles}}, \bibinfo
  {author} {\bibfnamefont {T.}~\bibnamefont {Lahaye}}, \bibinfo {author}
  {\bibfnamefont {A.}~\bibnamefont {Browaeys}}, \bibinfo {author}
  {\bibfnamefont {G.-O.}\ \bibnamefont {Reymond}}, \ and\ \bibinfo {author}
  {\bibfnamefont {C.}~\bibnamefont {Jurczak}},\ }\href {\doibase
  10.22331/q-2020-09-21-327} {\bibfield  {journal} {\bibinfo  {journal}
  {Quantum}\ }\textbf {\bibinfo {volume} {4}},\ \bibinfo {pages} {327}
  (\bibinfo {year} {2020})}\BibitemShut {NoStop}%
\bibitem [{\citenamefont {Ebadi}\ \emph {et~al.}(2021)\citenamefont {Ebadi},
  \citenamefont {Wang}, \citenamefont {Levine}, \citenamefont {Keesling},
  \citenamefont {Semeghini}, \citenamefont {Omran}, \citenamefont {Bluvstein},
  \citenamefont {Samajdar}, \citenamefont {Pichler}, \citenamefont {Ho},
  \citenamefont {Choi}, \citenamefont {Sachdev}, \citenamefont {Greiner},
  \citenamefont {Vuleti{\'c}},\ and\ \citenamefont
  {Lukin}}]{Ebadi2021QuantumPO}%
  \BibitemOpen
  \bibfield  {author} {\bibinfo {author} {\bibfnamefont {S.}~\bibnamefont
  {Ebadi}}, \bibinfo {author} {\bibfnamefont {T.~T.}\ \bibnamefont {Wang}},
  \bibinfo {author} {\bibfnamefont {H.}~\bibnamefont {Levine}}, \bibinfo
  {author} {\bibfnamefont {A.}~\bibnamefont {Keesling}}, \bibinfo {author}
  {\bibfnamefont {G.}~\bibnamefont {Semeghini}}, \bibinfo {author}
  {\bibfnamefont {A.}~\bibnamefont {Omran}}, \bibinfo {author} {\bibfnamefont
  {D.}~\bibnamefont {Bluvstein}}, \bibinfo {author} {\bibfnamefont
  {R.}~\bibnamefont {Samajdar}}, \bibinfo {author} {\bibfnamefont
  {H.}~\bibnamefont {Pichler}}, \bibinfo {author} {\bibfnamefont {W.~W.}\
  \bibnamefont {Ho}}, \bibinfo {author} {\bibfnamefont {S.}~\bibnamefont
  {Choi}}, \bibinfo {author} {\bibfnamefont {S.}~\bibnamefont {Sachdev}},
  \bibinfo {author} {\bibfnamefont {M.}~\bibnamefont {Greiner}}, \bibinfo
  {author} {\bibfnamefont {V.}~\bibnamefont {Vuleti{\'c}}}, \ and\ \bibinfo
  {author} {\bibfnamefont {M.~D.}\ \bibnamefont {Lukin}},\ }\href@noop {}
  {\bibfield  {journal} {\bibinfo  {journal} {Nature}\ }\textbf {\bibinfo
  {volume} {595 7866}},\ \bibinfo {pages} {227} (\bibinfo {year}
  {2021})}\BibitemShut {NoStop}%
\bibitem [{\citenamefont {Wu}\ \emph {et~al.}(2019)\citenamefont {Wu},
  \citenamefont {Kumar}, \citenamefont {Giraldo},\ and\ \citenamefont
  {Weiss}}]{Wu_2019}%
  \BibitemOpen
  \bibfield  {author} {\bibinfo {author} {\bibfnamefont {T.-Y.}\ \bibnamefont
  {Wu}}, \bibinfo {author} {\bibfnamefont {A.}~\bibnamefont {Kumar}}, \bibinfo
  {author} {\bibfnamefont {F.}~\bibnamefont {Giraldo}}, \ and\ \bibinfo
  {author} {\bibfnamefont {D.~S.}\ \bibnamefont {Weiss}},\ }\href {\doibase
  10.1038/s41567-019-0478-8} {\bibfield  {journal} {\bibinfo  {journal} {Nature
  Physics}\ }\textbf {\bibinfo {volume} {15}},\ \bibinfo {pages} {538}
  (\bibinfo {year} {2019})}\BibitemShut {NoStop}%
\bibitem [{\citenamefont {Graham}\ \emph {et~al.}(2019)\citenamefont {Graham},
  \citenamefont {Kwon}, \citenamefont {Grinkemeyer}, \citenamefont {Marra},
  \citenamefont {Jiang}, \citenamefont {Lichtman}, \citenamefont {Sun},
  \citenamefont {Ebert},\ and\ \citenamefont
  {Saffman}}]{PhysRevLett.123.230501}%
  \BibitemOpen
  \bibfield  {author} {\bibinfo {author} {\bibfnamefont {T.~M.}\ \bibnamefont
  {Graham}}, \bibinfo {author} {\bibfnamefont {M.}~\bibnamefont {Kwon}},
  \bibinfo {author} {\bibfnamefont {B.}~\bibnamefont {Grinkemeyer}}, \bibinfo
  {author} {\bibfnamefont {Z.}~\bibnamefont {Marra}}, \bibinfo {author}
  {\bibfnamefont {X.}~\bibnamefont {Jiang}}, \bibinfo {author} {\bibfnamefont
  {M.~T.}\ \bibnamefont {Lichtman}}, \bibinfo {author} {\bibfnamefont
  {Y.}~\bibnamefont {Sun}}, \bibinfo {author} {\bibfnamefont {M.}~\bibnamefont
  {Ebert}}, \ and\ \bibinfo {author} {\bibfnamefont {M.}~\bibnamefont
  {Saffman}},\ }\href {\doibase 10.1103/PhysRevLett.123.230501} {\bibfield
  {journal} {\bibinfo  {journal} {Phys. Rev. Lett.}\ }\textbf {\bibinfo
  {volume} {123}},\ \bibinfo {pages} {230501} (\bibinfo {year}
  {2019})}\BibitemShut {NoStop}%
\bibitem [{\citenamefont {Brennen}\ \emph {et~al.}(1999)\citenamefont
  {Brennen}, \citenamefont {Caves}, \citenamefont {Jessen},\ and\ \citenamefont
  {Deutsch}}]{PhysRevLett.82.1060}%
  \BibitemOpen
  \bibfield  {author} {\bibinfo {author} {\bibfnamefont {G.~K.}\ \bibnamefont
  {Brennen}}, \bibinfo {author} {\bibfnamefont {C.~M.}\ \bibnamefont {Caves}},
  \bibinfo {author} {\bibfnamefont {P.~S.}\ \bibnamefont {Jessen}}, \ and\
  \bibinfo {author} {\bibfnamefont {I.~H.}\ \bibnamefont {Deutsch}},\ }\href
  {\doibase 10.1103/PhysRevLett.82.1060} {\bibfield  {journal} {\bibinfo
  {journal} {Phys. Rev. Lett.}\ }\textbf {\bibinfo {volume} {82}},\ \bibinfo
  {pages} {1060} (\bibinfo {year} {1999})}\BibitemShut {NoStop}%
\bibitem [{\citenamefont {Jaksch}\ \emph {et~al.}(1999)\citenamefont {Jaksch},
  \citenamefont {Briegel}, \citenamefont {Cirac}, \citenamefont {Gardiner},\
  and\ \citenamefont {Zoller}}]{PhysRevLett.82.1975}%
  \BibitemOpen
  \bibfield  {author} {\bibinfo {author} {\bibfnamefont {D.}~\bibnamefont
  {Jaksch}}, \bibinfo {author} {\bibfnamefont {H.-J.}\ \bibnamefont {Briegel}},
  \bibinfo {author} {\bibfnamefont {J.~I.}\ \bibnamefont {Cirac}}, \bibinfo
  {author} {\bibfnamefont {C.~W.}\ \bibnamefont {Gardiner}}, \ and\ \bibinfo
  {author} {\bibfnamefont {P.}~\bibnamefont {Zoller}},\ }\href {\doibase
  10.1103/PhysRevLett.82.1975} {\bibfield  {journal} {\bibinfo  {journal}
  {Phys. Rev. Lett.}\ }\textbf {\bibinfo {volume} {82}},\ \bibinfo {pages}
  {1975} (\bibinfo {year} {1999})}\BibitemShut {NoStop}%
\bibitem [{\citenamefont {You}\ and\ \citenamefont
  {Chapman}(2000)}]{PhysRevA.62.052302}%
  \BibitemOpen
  \bibfield  {author} {\bibinfo {author} {\bibfnamefont {L.}~\bibnamefont
  {You}}\ and\ \bibinfo {author} {\bibfnamefont {M.~S.}\ \bibnamefont
  {Chapman}},\ }\href {\doibase 10.1103/PhysRevA.62.052302} {\bibfield
  {journal} {\bibinfo  {journal} {Phys. Rev. A}\ }\textbf {\bibinfo {volume}
  {62}},\ \bibinfo {pages} {052302} (\bibinfo {year} {2000})}\BibitemShut
  {NoStop}%
\bibitem [{\citenamefont {Shi}(2022)}]{Shi_2022}%
  \BibitemOpen
  \bibfield  {author} {\bibinfo {author} {\bibfnamefont {X.-F.}\ \bibnamefont
  {Shi}},\ }\href {\doibase 10.1088/2058-9565/ac18b8} {\bibfield  {journal}
  {\bibinfo  {journal} {Quantum Science and Technology}\ }\textbf {\bibinfo
  {volume} {7}},\ \bibinfo {pages} {023002} (\bibinfo {year}
  {2022})}\BibitemShut {NoStop}%
\bibitem [{\citenamefont {Derevianko}\ \emph {et~al.}(2015)\citenamefont
  {Derevianko}, \citenamefont {K\'om\'ar}, \citenamefont {Topcu}, \citenamefont
  {Kroeze},\ and\ \citenamefont {Lukin}}]{Dereviyanko2015}%
  \BibitemOpen
  \bibfield  {author} {\bibinfo {author} {\bibfnamefont {A.}~\bibnamefont
  {Derevianko}}, \bibinfo {author} {\bibfnamefont {P.}~\bibnamefont
  {K\'om\'ar}}, \bibinfo {author} {\bibfnamefont {T.}~\bibnamefont {Topcu}},
  \bibinfo {author} {\bibfnamefont {R.~M.}\ \bibnamefont {Kroeze}}, \ and\
  \bibinfo {author} {\bibfnamefont {M.~D.}\ \bibnamefont {Lukin}},\ }\href
  {\doibase 10.1103/PhysRevA.92.063419} {\bibfield  {journal} {\bibinfo
  {journal} {Phys. Rev. A}\ }\textbf {\bibinfo {volume} {92}},\ \bibinfo
  {pages} {063419} (\bibinfo {year} {2015})}\BibitemShut {NoStop}%
\bibitem [{\citenamefont {Keating}\ \emph {et~al.}(2013)\citenamefont
  {Keating}, \citenamefont {Goyal}, \citenamefont {Jau}, \citenamefont
  {Biedermann}, \citenamefont {Landahl},\ and\ \citenamefont
  {Deutsch}}]{Deutsch2013}%
  \BibitemOpen
  \bibfield  {author} {\bibinfo {author} {\bibfnamefont {T.}~\bibnamefont
  {Keating}}, \bibinfo {author} {\bibfnamefont {K.}~\bibnamefont {Goyal}},
  \bibinfo {author} {\bibfnamefont {Y.-Y.}\ \bibnamefont {Jau}}, \bibinfo
  {author} {\bibfnamefont {G.~W.}\ \bibnamefont {Biedermann}}, \bibinfo
  {author} {\bibfnamefont {A.~J.}\ \bibnamefont {Landahl}}, \ and\ \bibinfo
  {author} {\bibfnamefont {I.~H.}\ \bibnamefont {Deutsch}},\ }\href {\doibase
  10.1103/PhysRevA.87.052314} {\bibfield  {journal} {\bibinfo  {journal} {Phys.
  Rev. A}\ }\textbf {\bibinfo {volume} {87}},\ \bibinfo {pages} {052314}
  (\bibinfo {year} {2013})}\BibitemShut {NoStop}%
\bibitem [{\citenamefont {de~L\'es\'eleuc}\ \emph {et~al.}(2018)\citenamefont
  {de~L\'es\'eleuc}, \citenamefont {Barredo}, \citenamefont {Lienhard},
  \citenamefont {Browaeys},\ and\ \citenamefont {Lahaye}}]{PhysRevA.97.053803}%
  \BibitemOpen
  \bibfield  {author} {\bibinfo {author} {\bibfnamefont {S.}~\bibnamefont
  {de~L\'es\'eleuc}}, \bibinfo {author} {\bibfnamefont {D.}~\bibnamefont
  {Barredo}}, \bibinfo {author} {\bibfnamefont {V.}~\bibnamefont {Lienhard}},
  \bibinfo {author} {\bibfnamefont {A.}~\bibnamefont {Browaeys}}, \ and\
  \bibinfo {author} {\bibfnamefont {T.}~\bibnamefont {Lahaye}},\ }\href
  {\doibase 10.1103/PhysRevA.97.053803} {\bibfield  {journal} {\bibinfo
  {journal} {Phys. Rev. A}\ }\textbf {\bibinfo {volume} {97}},\ \bibinfo
  {pages} {053803} (\bibinfo {year} {2018})}\BibitemShut {NoStop}%
\bibitem [{\citenamefont {Levine}\ \emph {et~al.}(2018)\citenamefont {Levine},
  \citenamefont {Keesling}, \citenamefont {Omran}, \citenamefont {Bernien},
  \citenamefont {Schwartz}, \citenamefont {Zibrov}, \citenamefont {Endres},
  \citenamefont {Greiner}, \citenamefont {Vuleti\ifmmode~\acute{c}\else
  \'{c}\fi{}},\ and\ \citenamefont {Lukin}}]{PhysRevLett.121.123603}%
  \BibitemOpen
  \bibfield  {author} {\bibinfo {author} {\bibfnamefont {H.}~\bibnamefont
  {Levine}}, \bibinfo {author} {\bibfnamefont {A.}~\bibnamefont {Keesling}},
  \bibinfo {author} {\bibfnamefont {A.}~\bibnamefont {Omran}}, \bibinfo
  {author} {\bibfnamefont {H.}~\bibnamefont {Bernien}}, \bibinfo {author}
  {\bibfnamefont {S.}~\bibnamefont {Schwartz}}, \bibinfo {author}
  {\bibfnamefont {A.~S.}\ \bibnamefont {Zibrov}}, \bibinfo {author}
  {\bibfnamefont {M.}~\bibnamefont {Endres}}, \bibinfo {author} {\bibfnamefont
  {M.}~\bibnamefont {Greiner}}, \bibinfo {author} {\bibfnamefont
  {V.}~\bibnamefont {Vuleti\ifmmode~\acute{c}\else \'{c}\fi{}}}, \ and\
  \bibinfo {author} {\bibfnamefont {M.~D.}\ \bibnamefont {Lukin}},\ }\href
  {\doibase 10.1103/PhysRevLett.121.123603} {\bibfield  {journal} {\bibinfo
  {journal} {Phys. Rev. Lett.}\ }\textbf {\bibinfo {volume} {121}},\ \bibinfo
  {pages} {123603} (\bibinfo {year} {2018})}\BibitemShut {NoStop}%
\bibitem [{\citenamefont {Levine}\ \emph {et~al.}(2019)\citenamefont {Levine},
  \citenamefont {Keesling}, \citenamefont {Semeghini}, \citenamefont {Omran},
  \citenamefont {Wang}, \citenamefont {Ebadi}, \citenamefont {Bernien},
  \citenamefont {Greiner}, \citenamefont {Vuleti\ifmmode~\acute{c}\else
  \'{c}\fi{}}, \citenamefont {Pichler},\ and\ \citenamefont
  {Lukin}}]{Lukin2019}%
  \BibitemOpen
  \bibfield  {author} {\bibinfo {author} {\bibfnamefont {H.}~\bibnamefont
  {Levine}}, \bibinfo {author} {\bibfnamefont {A.}~\bibnamefont {Keesling}},
  \bibinfo {author} {\bibfnamefont {G.}~\bibnamefont {Semeghini}}, \bibinfo
  {author} {\bibfnamefont {A.}~\bibnamefont {Omran}}, \bibinfo {author}
  {\bibfnamefont {T.~T.}\ \bibnamefont {Wang}}, \bibinfo {author}
  {\bibfnamefont {S.}~\bibnamefont {Ebadi}}, \bibinfo {author} {\bibfnamefont
  {H.}~\bibnamefont {Bernien}}, \bibinfo {author} {\bibfnamefont
  {M.}~\bibnamefont {Greiner}}, \bibinfo {author} {\bibfnamefont
  {V.}~\bibnamefont {Vuleti\ifmmode~\acute{c}\else \'{c}\fi{}}}, \bibinfo
  {author} {\bibfnamefont {H.}~\bibnamefont {Pichler}}, \ and\ \bibinfo
  {author} {\bibfnamefont {M.~D.}\ \bibnamefont {Lukin}},\ }\href {\doibase
  10.1103/PhysRevLett.123.170503} {\bibfield  {journal} {\bibinfo  {journal}
  {Phys. Rev. Lett.}\ }\textbf {\bibinfo {volume} {123}},\ \bibinfo {pages}
  {170503} (\bibinfo {year} {2019})}\BibitemShut {NoStop}%
\bibitem [{\citenamefont {Barredo}\ \emph {et~al.}(2014)\citenamefont
  {Barredo}, \citenamefont {Ravets}, \citenamefont {H.~Labuhn}, \citenamefont
  {Vernier}, \citenamefont {Nogrette}, \citenamefont {Lahaye},\ and\
  \citenamefont {Browaeys}}]{Browaeys2014}%
  \BibitemOpen
  \bibfield  {author} {\bibinfo {author} {\bibfnamefont {D.}~\bibnamefont
  {Barredo}}, \bibinfo {author} {\bibfnamefont {S.}~\bibnamefont {Ravets}},
  \bibinfo {author} {\bibfnamefont {L.~B.}\ \bibnamefont {H.~Labuhn}}, \bibinfo
  {author} {\bibfnamefont {A.}~\bibnamefont {Vernier}}, \bibinfo {author}
  {\bibfnamefont {F.}~\bibnamefont {Nogrette}}, \bibinfo {author}
  {\bibfnamefont {T.}~\bibnamefont {Lahaye}}, \ and\ \bibinfo {author}
  {\bibfnamefont {A.}~\bibnamefont {Browaeys}},\ }\href {\doibase
  10.1103/PhysRevLett.112.183002} {\bibfield  {journal} {\bibinfo  {journal}
  {Phys. Rev. Lett.}\ }\textbf {\bibinfo {volume} {112}},\ \bibinfo {pages}
  {183002} (\bibinfo {year} {2014})}\BibitemShut {NoStop}%
\bibitem [{\citenamefont {Ravets}\ \emph {et~al.}(2015)\citenamefont {Ravets},
  \citenamefont {Labuhn}, \citenamefont {Barredo}, \citenamefont {Lahaye},\
  and\ \citenamefont {Browaeys}}]{Browaeys2015}%
  \BibitemOpen
  \bibfield  {author} {\bibinfo {author} {\bibfnamefont {S.}~\bibnamefont
  {Ravets}}, \bibinfo {author} {\bibfnamefont {H.}~\bibnamefont {Labuhn}},
  \bibinfo {author} {\bibfnamefont {D.}~\bibnamefont {Barredo}}, \bibinfo
  {author} {\bibfnamefont {T.}~\bibnamefont {Lahaye}}, \ and\ \bibinfo {author}
  {\bibfnamefont {A.}~\bibnamefont {Browaeys}},\ }\href {\doibase
  10.1103/PhysRevA.92.020701} {\bibfield  {journal} {\bibinfo  {journal} {Phys.
  Rev. A}\ }\textbf {\bibinfo {volume} {92}},\ \bibinfo {pages} {020701(R)}
  (\bibinfo {year} {2015})}\BibitemShut {NoStop}%
\bibitem [{\citenamefont {S\o{}rensen}\ and\ \citenamefont
  {M\o{}lmer}(1999)}]{Sorensen1999}%
  \BibitemOpen
  \bibfield  {author} {\bibinfo {author} {\bibfnamefont {A.}~\bibnamefont
  {S\o{}rensen}}\ and\ \bibinfo {author} {\bibfnamefont {K.}~\bibnamefont
  {M\o{}lmer}},\ }\href {\doibase 10.1103/PhysRevLett.82.1971} {\bibfield
  {journal} {\bibinfo  {journal} {Phys. Rev. Lett.}\ }\textbf {\bibinfo
  {volume} {82}},\ \bibinfo {pages} {1971} (\bibinfo {year}
  {1999})}\BibitemShut {NoStop}%
\bibitem [{\citenamefont {Walker}\ and\ \citenamefont
  {Saffman}(2008)}]{Walker_2008}%
  \BibitemOpen
  \bibfield  {author} {\bibinfo {author} {\bibfnamefont {T.~G.}\ \bibnamefont
  {Walker}}\ and\ \bibinfo {author} {\bibfnamefont {M.}~\bibnamefont
  {Saffman}},\ }\href {\doibase 10.1103/physreva.77.032723} {\bibfield
  {journal} {\bibinfo  {journal} {Physical Review A}\ }\textbf {\bibinfo
  {volume} {77}} (\bibinfo {year} {2008}),\
  10.1103/physreva.77.032723}\BibitemShut {NoStop}%
\bibitem [{\citenamefont {Happer}(1972)}]{HapperReview}%
  \BibitemOpen
  \bibfield  {author} {\bibinfo {author} {\bibfnamefont {W.}~\bibnamefont
  {Happer}},\ }\href {\doibase 10.1103/RevModPhys.44.169} {\bibfield  {journal}
  {\bibinfo  {journal} {Rev. Mod. Phys.}\ }\textbf {\bibinfo {volume} {44}},\
  \bibinfo {pages} {169} (\bibinfo {year} {1972})}\BibitemShut {NoStop}%
\bibitem [{\citenamefont {Omont}(1977)}]{OmontReview}%
  \BibitemOpen
  \bibfield  {author} {\bibinfo {author} {\bibfnamefont {A.}~\bibnamefont
  {Omont}},\ }\href {\doibase https://doi.org/10.1016/0079-6727(79)90003-X}
  {\bibfield  {journal} {\bibinfo  {journal} {Progress in Quantum Electronics}\
  }\textbf {\bibinfo {volume} {5}},\ \bibinfo {pages} {69} (\bibinfo {year}
  {1977})}\BibitemShut {NoStop}%
\bibitem [{\citenamefont {Varshalovich}\ \emph {et~al.}(1988)\citenamefont
  {Varshalovich}, \citenamefont {Moskalev},\ and\ \citenamefont
  {Khersonskii}}]{Varshalovich}%
  \BibitemOpen
  \bibfield  {author} {\bibinfo {author} {\bibfnamefont {D.~A.}\ \bibnamefont
  {Varshalovich}}, \bibinfo {author} {\bibfnamefont {A.~N.}\ \bibnamefont
  {Moskalev}}, \ and\ \bibinfo {author} {\bibfnamefont {V.~K.}\ \bibnamefont
  {Khersonskii}},\ }\href@noop {} {\emph {\bibinfo {title} {Quantum theory of
  angular momentum}}}\ (\bibinfo  {publisher} {World Scientific},\ \bibinfo
  {year} {1988})\BibitemShut {NoStop}%
\bibitem [{\citenamefont {Messiah}(1961)}]{messiah61}%
  \BibitemOpen
  \bibfield  {author} {\bibinfo {author} {\bibfnamefont {A.}~\bibnamefont
  {Messiah}},\ }\href@noop {} {\emph {\bibinfo {title} {Quantum Mechanics
  Volume II}}}\ (\bibinfo  {publisher} {Elsevier Science B.V.},\ \bibinfo
  {year} {1961})\BibitemShut {NoStop}%
\bibitem [{\citenamefont {Landau}\ and\ \citenamefont
  {Lifshitz}(1977)}]{LaLfIII}%
  \BibitemOpen
  \bibfield  {author} {\bibinfo {author} {\bibfnamefont {L.}~\bibnamefont
  {Landau}}\ and\ \bibinfo {author} {\bibfnamefont {E.}~\bibnamefont
  {Lifshitz}},\ }\href@noop {} {\emph {\bibinfo {title} {Quantum Mechanics}}},\
  \bibinfo {edition} {3rd}\ ed.,\ \bibinfo {series} {Course of Theoretical
  Physics}, Vol.~\bibinfo {volume} {3}\ (\bibinfo  {publisher} {Pergamon},\
  \bibinfo {year} {1977})\BibitemShut {NoStop}%
\bibitem [{\citenamefont {Gerasimov}\ \emph {et~al.}(2022)\citenamefont
  {Gerasimov}, \citenamefont {Yusupov}, \citenamefont {Moiseevsky},
  \citenamefont {Vyborny}, \citenamefont {Tikhonov}, \citenamefont {Kulik},
  \citenamefont {Straupe}, \citenamefont {Sukenik},\ and\ \citenamefont
  {Kupriyanov}}]{TwoQubitsPaper2022}%
  \BibitemOpen
  \bibfield  {author} {\bibinfo {author} {\bibfnamefont {L.}~\bibnamefont
  {Gerasimov}}, \bibinfo {author} {\bibfnamefont {R.}~\bibnamefont {Yusupov}},
  \bibinfo {author} {\bibfnamefont {A.}~\bibnamefont {Moiseevsky}}, \bibinfo
  {author} {\bibfnamefont {I.}~\bibnamefont {Vyborny}}, \bibinfo {author}
  {\bibfnamefont {K.}~\bibnamefont {Tikhonov}}, \bibinfo {author}
  {\bibfnamefont {S.}~\bibnamefont {Kulik}}, \bibinfo {author} {\bibfnamefont
  {S.}~\bibnamefont {Straupe}}, \bibinfo {author} {\bibfnamefont
  {C.}~\bibnamefont {Sukenik}}, \ and\ \bibinfo {author} {\bibfnamefont
  {D.}~\bibnamefont {Kupriyanov}},\ }\href {https://arxiv.org/abs/2205.03383}
  {\bibfield  {journal} {\bibinfo  {journal} {arxiv.org/abs/2205.03383}\ }
  (\bibinfo {year} {2022})}\BibitemShut {NoStop}%
\end{thebibliography}%

\end{document}